\documentstyle[preprint,aps,epsf,floats]{revtex}

\begin{document}

\draft

\tighten

\preprint{\vbox{\hfill hep-ph/0005174 \\
          \vbox{\hfill May 2000} \\
          \vbox{\hfill revised July 2000} \\
          \vbox{\vskip0.5in}
         }}

\title{Relic Abundances and the Boltzmann Equation}

\author{Mark Srednicki\footnote{E--mail: mark@vulcan.physics.ucsb.edu}}

\address{Department of Physics, University of California,
         Santa Barbara, CA 93106 }

\maketitle

\begin{abstract}
\normalsize{
I discuss the validity of the quantum Boltzmann equation for the 
calculation of WIMP relic densities. 
}
\end{abstract}

\vskip1in

\centerline{\em Talk given at DM2000,} 
\centerline{\em Sources and Detection of Dark Matter/Energy in the Universe,}
\centerline{\em Marina del Rey, CA, February 23--25, 2000}

\pacs{}

The quantum Boltzmann equation (QBE) is the starting point for the calculation
of the relic density of stable, weakly interacting massive particles (WIMPs)
that were once in thermal equilibrium.  This describes a number of dark matter
candidates (such as the lightest supersymmetric particle, or LSP).
A linearized form of the QBE is also used to calculate the power spectrum
of fluctuations in the cosmic microwave background.  We rely on the
results of these calculations, and expect them to be accurate to
a high degree of precision.

It is therefore somewhat disconcerting to realize that the theoretical
derivation of the QBE is not at all straightforward, and that there can be 
important corrections.  
(For a good general discussion of the key issues in the simpler
context of nonrelativistic quantum mechanics, see \cite{rm}.)
Furthermore, I am unaware of any experimental verification of the
QBE in which the relevant cross sections have been calculated from first
principles, with no phenomenological input.  Thus astrophysics leads us
to a new frontier for the QBE.  

In a series of papers, Matsumoto and Yoshimura (MY) \cite{my}
have carefully reconsidered the derivation of the QBE for WIMP relic 
densities.  They found a surprising result: a previously unsuspected
correction term that becomes dominant at low temperatures, and
changes the standard results.  Furthermore, this correction can
be computed and understood using the methods of thermal field theory;
the subtleties in the derivation of the QBE do not enter into it.

Simply stated, the result of MY is that the number density
$n$ of a gas of WIMPS, each with mass $M$, in thermal equilibrium
at a temperature $T \ll M$ with a gas of massless particles, is given by
\begin{equation}
n = (2\pi)^{-3/2} (M T)^{3/2} e^{-M/T} + c\lambda^2 T^6/M^3 + \ldots\;,
\label{n}
\end{equation}
where $\lambda$ is a coupling constant and $c$ is a numerical factor.
The first term represents the usual result for noninteracting particles,
and carries a Boltzmann suppression factor $e^{-M/T}$; the second term
is a loop correction, and it does not have this suppression.  Therefore,
no matter how small $\lambda$ is, at sufficiently low temperatures the
second term will dominate.   

Is this result correct?  We must first ask for the definition of $n$.
MY use a model of a heavy spin-zero boson (represented by a real scalar 
field $\varphi$) interacting with a massless spin-zero boson (represented
by a real scalar field $\chi$).  The scalar potential is 
\begin{equation}
V(\varphi,\chi)=  {\textstyle{1\over2}}M^2\varphi^2
                 +{\textstyle{1\over24}}\lambda_\varphi\varphi^4
                 +{\textstyle{1\over24}}\lambda_\chi\chi^4
                 +{\textstyle{1\over4}}\lambda \varphi^2\chi^2.
\label{v}
\end{equation}
We assume that 
$\lambda_\varphi \ll \lambda^2$,
$\lambda \ll \lambda_\chi$, and
$\lambda_\chi < 1$.
This hierarchy among the couplings allows the light $\chi$ particles to 
function as an efficient heat bath for the heavy $\varphi$ particles.
Then, for $T\ll M$, MY define the number density $n$
of heavy particles via
\begin{equation}
n = {1\over M}\langle{\cal H}_\varphi\rangle 
  = {1\over M}\left[{\mathop{\rm Tr} {\cal H}_\varphi e^{-H/T}
                    \over
                    \mathop{\rm Tr} e^{-H/T} }
                    - \langle 0|{\cal H}_\varphi|0\rangle\right],
\label{n2}
\end{equation}
where $H$ is the total hamiltonian, and
\begin{equation}
{\cal H}_{\varphi} = {\textstyle{1\over2}}\dot\varphi^2 +
                     {\textstyle{1\over2}}(\nabla\varphi)^2 + 
                     {\textstyle{1\over2}}M^2\varphi^2 +
                     \hbox{counterterms}
\label{hphi}
\end{equation}
is the free-field part of the $\varphi$ hamiltonian,
plus counterterms (some of which involve the $\chi$ field)
that are necessary to remove infinities in this composite operator.

Eq.~(\ref{n2}) is a highly plausible definition of the heavy particle
number density (at low temperatures).
However, it does not correspond in any obvious way to how 
this number density would be determined experimentally.  
Standard methods all involve a search for individual, 
on-shell $\varphi$ particles.
Real-world examples of this include present-day
dark matter searches, and measurements of the
cosmic microwave background radiation.  
Therefore, it is possible to suspect that the definition
(\ref{n2}) is not appropriate, and this is the reason for
the surprising lack of Boltzmann suppression in the loop corrections
to $n$.

In \cite{as}, this issue was investigated using a simple Caldeira--Leggett
model \cite{cl}.  In this exactly solvable model, 
a ``system'' (represented by a harmonic oscillator) 
is coupled to an ``environment'' (represented by more oscillators)
via an ``interaction''.
We found that, at low temperatures, the ``interaction'' energy was
always comparable to the ``system'' energy, making the identification
of the ``system'' problematic.  

Consider now a slight variation of the MY model, in which the heavy
field is complex, and carries a conserved U(1) charge,
while the light field remains real and neutral \cite{ms}.
In this model, we can study the charge fluctuations in a given volume $V$.
For noninteracting particles, Poisson statistics for the number of
positively and negatively charged particles in $V$ results in
\begin{equation}
\langle Q^2\rangle = n V.
\label{q2}
\end{equation}
We see that the charge fluctuations give us a measurement of the
total number of heavy particles in a given volume.  If $Q$ represents
electric charge, we can in principle measure
$\langle Q^2\rangle$ without tracking individual heavy particles.
Furthermore, it seems highly unlikely that weak interactions could
significantly modify eq.~(\ref{q2}).  If $\langle Q^2\rangle$ 
is either much larger or much smaller than $nV$,
then the movements of positive and negative particles
into and out of $V$ would have to be
highly correlated (in order to suppress or enhance the charge 
fluctuations in $V$).  This is inconsistent 
with the usual notion of a gas of particles that move freely and
independently between occasional scatterings, 
and would appear to require strong interactions.

We therefore {\it define\/} the number density of heavy particles, 
for $T\ll M$, via eq.~(\ref{q2}).  We can now compute the $O(\lambda^2)$
loop corrections to $\langle Q^2\rangle$,
and see whether or not they are Boltzmann suppressed.
The answer is: they are.  Details of the loop graph
evaluation are given in \cite{ms}, but there is a simple
argument for the Boltzmann suppression of $\langle Q^2\rangle$
to all orders in perturbation theory.
Consider the partition function 
\begin{equation}
Z=\mathop{\rm Tr}e^{-(H - \mu Q)/T}
\label{z}
\end{equation}
where $\mu$ is a chemical potential for the charge $Q$.
We then have
\begin{equation}
\langle Q^2\rangle = T^2\,{\partial^2\over\partial\mu^2}\ln Z
                        \biggr|_{\mu=0}.
\label{q22}
\end{equation}
We will now show that all $\mu$-dependent terms in $Z$ are Boltzmann
suppressed.  We write
\begin{equation}
Z=\sum_\alpha e^{-(E_\alpha-\mu Q_\alpha)/T},
\label{z2}
\end{equation}
where the sum is over a basis of energy and charge eigenstates.
States that yield a $\mu$-dependent contribution to $Z$ must 
have $Q\ne0$.  However, in a weakly coupled theory,
the lowest-energy state with $Q\ne0$
is a state consisting of a single zero-momentum $\varphi$ particle.
This is an exact energy eigenstate with energy $M$; 
its contribution to $Z$ is obviously Boltzmann suppressed.  
Since all other states with $Q\ne0$ have higher energy, their contributions
to $Z$ are Boltzmann suppressed as well.
Eq.~(\ref{q22}) then implies that $\langle Q^2\rangle$ is 
Boltzmann suppressed.  

In the original model of MY, there is no conserved charge
whose fluctuations can be measured to determine the number density $n$
of heavy particles.  However, is seems likely that any definition of $n$
that corresponds to an experimental measurement will yield a
result that is Boltzmann suppressed.  Assuming this, a  
simple mathematical definition of $n$ is 
$n=(1/M)\rho_{\rm\scriptscriptstyle BS}$,
where
$\rho_{\rm\scriptscriptstyle BS}$
represents those terms in the total energy density that
have a Boltzmann-suppression factor.  
Any unsuppressed terms in $\rho$ are to be thought of
as corrections to the energy density of the gas of light
particles.  This interpretation is in fact the usual
one in quantum electrodynamics \cite{qed}, where unsuppressed loop corrections
to the total energy density $\rho$ (at temperatures
well below the electron mass) are attributed to self-interactions
among photons that arise after integrating 
the electron-positron field out of the functional integral.
Recently, this point of view has been investigated in detail by 
Braaten and Jia \cite{bj} for the original MY model.  
They show that, after integrating out the
heavy $\varphi$ field to get an effective theory of the light $\chi$ field
alone, the $T^6/M^2$ term in $\rho_\varphi$ is obtained as the thermal 
expectation value of a particular higher-dimensional operator in the 
hamiltonian of the effective theory.  However, the coefficient of this
operator can be changed to any value (including zero) by a field redefinition.
This shows that no physical significance can be attached to the $T^6/M^2$ term
in $\rho_\varphi$.
   
We therefore conclude that the QBE for WIMP relic densities should not 
be corrected in the manner proposed by MY.  However, we should keep
in mind that the theoretical situation is not as clear as we might
like it to be, and so be on the lookout for possible further surprises.

\begin{acknowledgments}

I thank Anupam Singh and Scott Thomas for discussions,
and Motohiko Yoshimura for correspondence.
I also thank Gary Steigman for bringing ref.~\cite{bj} to my belated
attention.  This work was supported in part
by the National Science Foundation through grant PHY--97--22022, 
and by the Institute of Geophysics and Planetary Physics through grant 920.

\end{acknowledgments}

\end{document}